\documentclass[a4paper,notitlepage,12pt]{article}
\usepackage{amsmath,amsthm,amsfonts,amssymb}
\usepackage{graphicx}
\usepackage{ifpdf}
\usepackage[mathscr]{eucal}
\usepackage{cancellation}

\begin{document}

\centerline{\Large\textbf{Chirikov criterion of resonance overlapping }}
\centerline{\Large\textbf{for the model of molecular dynamics}}
\vskip2ex
\centerline{\Large\textbf{Guzev M.A.}}
\centerline{\small Institute for Applied Mathematics FEB RAS, Vladivostok}
\vskip1ex
\centerline{\small e-mail: guzev@iam.dvo.ru}
\vskip4ex
\dA=\hsize\hsize=8cm\dB=\dA\advance\dB-\hsize\divide\dB2
\setbox\wboxa=\vbox to 0pt{\footnotesize\vskip-.6\baselineskip\noindent  molecular dynamics, Lennard-Jones potential,
Chirikov criterion, chaotic dynamics\vss}
\hsize=12cm\dB=\dA\advance\dB-\hsize\divide\dB2
\setbox\wbox=\vbox{\footnotesize
\noindent The chaotic dynamics in a cell of particles' chain interacting by means of Lennard-Jones potential is
considered. Chirikov criterion of resonance overlapping  is used as the condition of chaos. The asymptotic
representation for this function at low and high energies is obtained for the function corresponding to the criterion.
\vskip.5ex

\textbf{Keywords:} \box\wboxa
\vskip\baselineskip
\textbf{Mathematical Subject Classification 2000: 37D45 n}}
\noindent\mbox{\kern\dB}\hbox{\box\wbox}
\hsize=\dA

{\subsection*{\large Introduction}}

It is well known [1] that investigation of the problem occurrence of chaotic dynamics naturally led to the need to
study simple models in which there is chaos. Further study of the real physical processes in which there is chaos
showed that many simple models reflect its main, for example, the occurrence of
stochastic separatrix layer, formation of nonlinear resonances and etc.

One-dimensional chains of particles have also been the subject of research in studying the problems of chaotic dynamics
[1] because some problems for them are reduced to the study of standard maps known for nonlinear dynamic systems. On
the other hand
investigation of the phenomena of chaos chain of particles is also important in connection with the active use of the
particle method for simulating the behavior of materials. Speaking about this method we consider the particles as point
masses, and not as discrete elements allowing to reduce the equations of continuum mechanics to the difference system
of ordinary differential equations. One of the most well-developed variants of the particle method is the method of
molecular dynamics in the classical version of which particles act as atoms and molecules. If interatomic potential is
known then the dynamics of molecular compounds can be modeled with high accuracy.

Despite the good correlation between the results of computer simulation observed
experiment and material behavior there is a problem of understanding their internal mechanisms in terms of nonlinear dynamics. For
one-dimensional system corresponding model problem is reduced to consideration of the particle mass $m$ in the cell. The
dynamics of particle is determined by the Hamiltonian
\begin{equation} \label{GrindEQ__1_}
H=H(p,q,t)=\frac{p^{2} }{2m} +V(q),\qquad V(q)=U(q)+U(\xi (t)-q),
\end{equation}
in which $p$ is the particle momentum, and form of the function $\xi (t)$ depends on the collective behavior of
particles in the system. Formula \eqref{GrindEQ__1_}
means that particle interacts with a stationary particle on the left side and interacts with a
particle moving in a law $q_{2} =\xi(t)$ on the right one. The interaction between the particles is characterized by
the potential $U(q)$.

The phenomenon of strong chaos for particles interacting by means of the Lennard-Jones potential $U(q)$ was considered
in [2]. Chirikov criterion of resonance overlapping [3] was used as the condition of chaos. The function  $K_{n,n-1}$
corresponding to the criterion was calculated numerically with respect to energy. In this article we obtain the
asymptotic representation for this function at low and high energies.
\vskip2ex

\subsection{\large Chirikov criterion of resonance overlapping}
Let us remind the model [2]. Interaction is determined by the Lennard-Jones potential
$U(q)=U_{0}\left[\left(\dfrac{a}{q}\right)^{\alpha}-\dfrac{\alpha}{\beta}\left(\dfrac{a}{q}\right)^{\beta}\right]$
with $\alpha>\beta>0$. The perturbation $\xi(t)$ is given by the formula $\xi(t)=2a\bigl(1+\varepsilon(t)\bigr),\;
\varepsilon(t)=\alpha _{1}\cos\omega _{1}t$ with $|\alpha _{1}|\ll1$. We introduce the dimensionless variables setting
$H\to H/U_{0}$, $q\to q/a-1$, $t\to t/t_{0}$, $\omega _{1} \to \omega _{1} t_{0} $  where $t_{0} =\sqrt{ma^{2} /U_{0} } $.
Then Hamiltonian \eqref{GrindEQ__1_} is written in the form
\begin{equation} \label{GrindEQ__2_}
H(p,q,t)=\frac{p^{2}}{2} +V(q),\qquad V(q)=U(q+1)+U(1+\varepsilon(t)-q).
\end{equation}
Given the small amplitude  $\varepsilon (t)$ we carry out the expansion of the Hamiltonian \eqref{GrindEQ__2_} with
respect to $\varepsilon (t)$:
\begin{equation} \label{GrindEQ__3_}
\begin{gathered}
H(p,q,t)=H_{0} (p,q)-\varepsilon (t)H_{1} (q)+\ldots,\\
H_0(p,q)=\frac{p^2}2{+}W(q),\; H_{1}(q)=\ddi Uq(1{-}q),\; W(q)=U(q{+}1){+}U(q{-}1).
\end{gathered}
\end{equation}
Action-angle variables $(I,\varphi )$ are commonly used in studying the dynamics of Hamiltonian systems [1]. The
unperturbed action is conserved along the phase trajectory and its value is determined by initial conditions $(p_{0}
,q_{0} )$. If you choose $p_{0} =0$ the correspondence between the unperturbed action $I$ and the parameter $q_{0} $ is
unique. Transformation to the action-angle variables can be carried out numerically in accordance with formulae
\begin{equation}
\mkern-20muI(q_{0})=\frac{2}{\pi}\int\limits_{0}^{q_{0}}p(q_{0},q)dq,\quad\varphi(q,q_{0})=\omega\int\limits_{q_{0}}^{q}\frac{dq}{p(q_{0},q)},
\quad\omega =\dfrac{\pi }{\displaystyle2\int\limits_{0}^{q_{0}}\dfrac{dq}{p(q_{0},q)}},
\end{equation}
where  $p(q_{0},q)=\sqrt{2\bigl(V(q_{0} )-V(q)\bigr)}$. We express the variables $(p,q)$  in the terms of
$(I,\varphi)$. Then Hamiltonian $H(p,q,t)$ can be expanded in a Fourier series:
\begin{equation} \label{GrindEQ__5_}\mkern-20mu
\begin{aligned}
H(p,q,t)&=H_{0}(I){-}\frac{1}{2}\alpha_{1}\sum_{n}H_{n}(I)[\cos(n\varphi{+}\varphi_{0}{-}\omega_{1}t){+}
\cos(n\varphi{+}\varphi_{0}{+}\omega_{1} t)],\\
H_{n}&=\frac{1}{\pi}\int\limits_{0}^{2\pi}d\varphi U^{(1)}(1{-}q)\cos n\varphi.
\end{aligned}
\end{equation}
Let us consider the resonance condition for a three of numbers $(n,I_n ,\omega _{1} )$:\;
$n\omega(I_n)=\omega _{1}.$

To analyze the dynamics of the system in the vicinity of nonlinear resonance we use the standard method [1]. In this
case the only resonant harmonic in \eqref{GrindEQ__5_} is fixed:
\begin{equation} \label{GrindEQ__6_}
H(p,q,t)=H_{0}(I)-\frac{1}{2}\alpha _{1}H_{n}(I)\cos(n\varphi{+}\varphi_{0}{-}\omega_{1} t).
\end{equation}
We introduce a new phase $\psi =n\varphi +\varphi _{0} -\omega _{1} t$ then the equations of motion corresponding to
Hamiltonian \eqref{GrindEQ__6_} have the form
\begin{equation} \label{GrindEQ__7_}
\begin{aligned}\ddi It&=-\Dpi H\psi=\frac12\alpha_{1}H_{n}(I)\sin\psi,\\
\ddi\psi t&=\Dpi HI=n\omega(I)-\omega_{1}-\frac12\alpha_{1}\cos\psi\ddi{}IH_{n}(I).
\end{aligned}
\end{equation}
As the equations \eqref{GrindEQ__7_} are examined in the vicinity of the resonance action $I_{n1} $ it is assumed that
the value $J=I{-}I_{n} $ is small. We expand $H_{0}(I)$, $\omega (I)$ and neglect the terms $J^{2}$
correspondingly. Then the equations \eqref{GrindEQ__7_} are written in the form:
$$\ddi Jt=\frac12\alpha_{1}H_{n}(I_{n})\sin\psi,\qquad\ddi\psi t=nJ\ddi\omega I(I_{n}).$$
The obtained equations correspond to the so-called nonlinear pendulum Hamiltonian [1]:
$$\bar{H}(J,\psi)=n\frac{J^{2}}{2}\ddi\omega I(I_{n})-\frac12\alpha_1H_{n}(I_{n})\cos\psi.$$
The width of the resonance $\Delta J$is calculated  from the condition:
$$
\frac{(\Delta J)^{2}}2\ddi\omega I(I_{n})=\frac12\alpha_{1}H_{n}(I_{n});\qquad
\Delta J=\sqrt{\raise1ex\hbox{$\alpha_{1}H_{n}(I_{n})$}\mskip-5mu\bigg/\ddi{\omega}I(I_{n})}.
$$
In terms of frequency $\omega $ we have
\begin{equation} \label{GrindEQ__8_}
\Delta\omega\approx\sqrt{\frac12\alpha_1H_{n}(I_{n})\ddi\omega I(I_{n})} .
\end{equation}

Consider the invariant curves for the resonant action  $I_n$, $I_{n-1} $. In order to form a chaotic region in
the plane $(p,q)$ it is necessary to destroy the curves corresponding to these values of the action. Let us assume that
$\Delta \Omega \;(\Delta I)$ is the corresponding width of the resonance and
$\delta \Omega=\omega (I_{n} )-\omega(I_{(n-1)} )$ ($\delta I{=}I_{n}{-}I_{(n-1)} $)
is the distance between resonances. The parameter $K$ characterizing the
degree of resonance overlapping is equal to $K{=}\Delta I\big/\delta I{\approx}\Delta \Omega\big /\delta \Omega $ [1]. In
accordance to Chirikov criterion [3] the overlap of resonance takes place on condition that $|K|\ge 1$ :
\begin{equation} \label{GrindEQ__9_}
K_{n,n-1} (q_{0} (\omega _{1} ))=\frac{\Delta \omega (I_{n} )+
\Delta \omega (I_{(n-1)} )}{\omega (I_{n} )-\omega (I_{(n-1)} )} \sim1.
\end{equation}
In [2] the function  $K_{3,2}$ was calculated numerically at different $|\alpha _{1} |\ll1$ with respect to $q_{0} $
(energy). In this article we construct an asymptotic representation for the functions
$K_{1,2}\bigl(q_{0}(\omega _{1})\bigr),\;K_{2,3}\bigl(q_{0}(\omega _{1})\bigr)$
at small and large energies.

\subsection{Approximate action-angle variables at small energy}
The criterion \eqref{GrindEQ__9_} includes the
frequency and its derivative \eqref{GrindEQ__8_}, and the value of the Fourier coefficient \eqref{GrindEQ__5_}. From
the point of view of physics the smallness of the energy values correspond to the fulfillment of the inequality
$E\ll1$. This allows us to use an approximation for $W(q)$ \eqref{GrindEQ__3_} as polynomial functions with respect to
$q$ and construct approximate formulae for the action-angle variables at $E\to 0$.
Approximation for the potential $W$ is obtained after expansion in the vicinity of the point $q=0$:
\begin{equation} \label{GrindEQ__10_}
W_{app} (q)=W_{0} +bq^{2} +cq^{4},\qquad b=U^{(2)} (1),\qquad c=\frac{U^{(4)}(1)}{12}.
\end{equation}
Let us  introduce the approximate action-angle variables assuming
$$J(E)=\frac{1}{\pi }\int\limits_{Q_{1} }^{Q_{2}}p_{app}dq_,\qquad\psi=\Omega t_{app},
\qquad t_{app}=\int\limits_{Q_{1} }^{q}\frac{dq}{p_{app} }$$
where $p_{app} =\sqrt{2\bigl(E-W_{app} (q)\bigr)}$, and the coordinates of points $Q_{i}$ are determined from the
equation $E=W_{app} (Q_{i} )$. The quantity  $\Omega =\Dpi EJ=2\dfrac\pi{T(E)}$ is  the cyclic
frequency where $T(E)$ is the period of motion in the potential \eqref{GrindEQ__10_}. Omitting simple calculations we
have the following presentation for $T(E)$:
$$
\begin{gathered}
\frac{T}{2}=\int\limits_{Q_{1}}^{Q_{2}}\frac{dq}{\sqrt{2\bigl(E-W_{app}(q)\bigr)}}=
\frac{1}{\sqrt{2c}}\int\limits_{-T_{1} }^{T_{1}}\frac{dl}{\sqrt{(T_{1} ^{2} -l^{2} )(A^{2} +l^{2} )}},\\
T_{1}=\sqrt{-b+\sqrt{b^{2}+\dfrac{4cH}{2b}}},\qquad H=E-W_{0},\qquad A^{2}=\frac{b+\sqrt{b^{2} +4cH} }{2c}.
\end{gathered}
$$
Substituting $t=B\cos\varphi $ we write this expression in the form :
\begin{equation} \label{GrindEQ__10a_}
\frac{T}{2}=\sqrt{\frac{2}{c}}\frac{1}{\sqrt{T_{1}^{2}+A^{2} }}\int\limits_{0}^{\pi/2}\frac{dt}{\sqrt{1-k^{2} \sin ^{2}\varphi}}=
\sqrt{\frac{2}{c}}\frac{1}{\sqrt{T_{1}^{2}+A^{2}}}{\bs K}(k),
\end{equation}
where ${\bs K}(k)$ is the complete elliptic integral. As a result obtain
\begin{equation} \label{GrindEQ__11_}
\psi=\frac{\pi }{2} \frac{F(\Phi ,k)}{{\bs K}(k)},\qquad q=T_{1} \cos\Phi
\end{equation}
where $F(\Phi,k)$ is the elliptic integral of the first order.

\subsection{Asymptotic behavior of different quantities at small energy}
We are interested in the asymptotic behavior of $T$ and the derivative $\Dpi TH{=}\Dpi TE$ at $H\to 0$, then
\begin{equation} \label{GrindEQ__12_}
k=\frac{\sqrt{Hc}}b\bigl(1+\underline{O}(H)\bigr),\qquad\frac{1}{\sqrt{T_{1}^{2}+A^{2}}}=
\sqrt{\frac{c}{b}} \left(1-\frac{Hb}{b^{2}}\right)+\underline{O}(H^{2}).
\end{equation}
The smallness of $k$ allows us to use the asymptotic formula for ${\bs K}(k)$:
\begin{equation} \label{GrindEQ__13_}
{\bs K}(k)=\frac{\pi }{2} \left\{1+\left(\frac{1}{2} \right)^{2} k^{2} +\left(\frac{1\cdot 3}{2\cdot 4} \right)^{2} k^{4} +\ldots\right\}.
\end{equation}
From here and \eqref{GrindEQ__10a_} we have
$$\frac{T}{2} =\frac{\pi }{\sqrt{2b} } \left(1-\frac{3Hc}{4b^{2} } \right)+\underline{O}(H^{2}).$$
Hence we immediately obtain the leading order in the expression for the frequency and its derivative which can be written as:
\begin{equation} \label{GrindEQ__14_}
\begin{aligned}\omega&=\sqrt{2b}+\underline{O}(H^{2})=\sqrt{2U^{(2)}(1)}+\underline{O}(H^{2}),\\
\Dpi\omega E&=\Dpi\omega H=\frac{3c}{\sqrt{2ba}}+\underline{O}(H)=\frac{U^{(4)}(1)}{4\sqrt{2U^{(2)}(1)}}+\underline{O}(H).
\end{aligned}
\end{equation}
Let us write the function $q(J,\psi )$  in  the terms of variables $(J,\psi )$ . The error does not exceed
$\underline{O}(H)$. Because of $k\sim \sqrt{H} $ \eqref{GrindEQ__12_} we use asymptotic formulae for $\mathbf K(k)$
\eqref{GrindEQ__13_} and ${\it F}(\Phi ,k)=\Phi +\ldots_{}^{} $ It results in $\psi =\Phi +\underline{O}(H)$ and  $\Phi
=\varphi +\underline{O}(H)$.

To calculate the Fourier coefficients \eqref{GrindEQ__5_} we expand the potential in the vicinity of $q=0$:
\begin{equation*}
H_{n}=\frac{1}{\pi }\int\limits_{0}^{2\pi }d\varphi\cos n\varphi\left[U^{(2)}(1)q+U^{(3)}(1)\frac{q^{2}}{2!}+
U^{(4)}(1)\frac{q^{3}}{3!}+\ldots\right] .
\end{equation*}
The function $q(J,\psi )$ is equal to $q=T_{1}\cos(\varphi{+}\underline{O}(H))$ where
$T_1=\sqrt{H/b}+\underline{O}(H^{2})$. Then we have
\begin{equation*}
H_{2}\sim H\frac{U^{(3)}(1)}{4U^{(2)}(1)},\qquad  H_{3} \sim \sqrt[{}]{H^{3} }.
\end{equation*}

From \eqref{GrindEQ__8_}, \eqref{GrindEQ__14_} we have inequalities
$|\Delta \omega (I_{1} )|\sim \sqrt[4]{H}\gg
|\Delta \omega (I_{2} )|\sim \sqrt{H} \gg\break|\Delta \omega (I_{3} )|\sim \sqrt[4]{H^{3} }$.
It allows us to obtain expression for functions $|K_{2,1} |,\;|K_{3,2} |$:
\begin{equation} \label{GrindEQ__17_}\mkern-20mu
\begin{aligned}
|K_{2,1} |&{\approx} \left|\frac{\Delta \omega (I_{1} )}{2(\omega (I_{2} )-\omega (I_{1} ))} \right|
{=}\left|\frac{\Delta \omega (I_{1} )}{\omega _{1} }\right|{\sim}
\frac1{\omega _{1}}\sqrt{\frac{1}{2}\alpha _{1} \sqrt{HU^{(2)} (1)} \frac{U^{(4)} (1)}{8U^{(2)} (1)} },\\
|K_{3,2} |&{\approx} \left|\frac{\Delta \omega (I_{2} )}{2(\omega (I_{3} )-\omega (I_{2} ))} \right|{=}
3\left|\frac{\Delta \omega (I_{2} )}{\omega _{1} }\right|{\sim}
\frac3{\omega _{1}}\sqrt{\frac{1}{2} \alpha _{1} H\frac{U^{(3)} (1)}{32U^{(2)} (1)} \frac{U^{(4)} (1)}{U^{(2)} (1)} }.
\end{aligned}
\end{equation}

\subsection{Approximate action-angle variables at high energy and asymptotic behavior of different quantities }
From the point of view of physics high values of energy $E$ correspond to the fulfillment of the inequality
$E/{U_{0}}{\gg}1$. Since the potential $W(q)=U(1{+}q)+U(1{-}q)$ is symmetric function with respect to $q\to -q$ it is enough to
consider it at the interval $(0,1)$. It is clear that for large values of $E$ turning point coordinate $q_{1}\sim1$.
Then the leading contribution in $V(q)$ is determined by $U(1-q)$. We introduce an approximate potential
\begin{equation} \label{GrindEQ__34_}
W_{big}(q)=\frac{1}{(1-q)^{\alpha }}
\end{equation}
and action-angle variables $J_{big} ,\phi $
\begin{equation} \label{GrindEQ__35_}
J_{big} (E)=\frac{2}{\pi }\int\limits_{Q_{1} }^{1}p_{big}dq,\qquad\phi =\Omega_{big}t_{big},\qquad
\Omega_{big}=\Dpi E{J_{big}},\qquad t_{big}=\int\limits_{Q_{1}}^{q}\frac{dq}{p_{big}},
\end{equation}
where function $p_{big}=\sqrt{2\bigl(E-W_{big} (q)\bigr)}$, and  the coordinate $Q_{1} $ is defined by the equation
\begin{equation} \label{GrindEQ__36_}
E=W_{big} (Q_{1} ).
\end{equation}
From a mathematical point of view, values $I$, $J_{big} $ determine the area in phase space for functions  $p$,
$p_{big} $  respectively. Area difference can be estimated by the sum of functions
\begin{equation} \label{GrindEQ__37_}
|I(E)-J_{big}(E)|\le \frac{2}{\pi }[|Q_{1} -q_{1}|\max |p|+|\int\limits_{Q_{1} }^{1}[p-p_{big}  ]dq|].
\end{equation}
Leading order of $Q_{1}{-}q_{1} $ is determined from \eqref{GrindEQ__34_}, \eqref{GrindEQ__36_} and is equal to
$Q_{1}{-}q_{1}=\dfrac{Q_{1}^{\alpha{-}\beta{+}1}}\beta\bigl(1+\underline{O}(Q_{1}^{\alpha{-}\beta})\bigr)$.
Since $\max|p|\le\sqrt{2E}$ then the first term in \eqref{GrindEQ__37_} has the following order
$(Q_{1} -q_{1} )\sqrt{E} \sim\dfrac{1}{E^{1/2+(1-\beta )/\alpha}}$. The integral term in \eqref{GrindEQ__37_} is equal to
$$\left|\int\limits_{\mskip4muQ_{1}}^{1}[p-p_{big}]dq\right|=
\left|\int\limits_{\mskip4muQ_{1}}^{1}\frac{W_{big}-W}{p+p_{big}}dq\right|\sim
\left|\int\limits_{\mskip4muQ_{1} }^{1}\frac{1}{p_{big}}dq\right|\sim
Q_{1}^{1-\beta +\alpha /2} \sim \frac{1}{E^{1/2+(1-\beta )/\alpha } }.$$
Hence
\begin{equation} \label{GrindEQ__38_}
I(E)-J_{big} (E)\sim \frac{1}{E^{1/2+(1-\beta )/\alpha } }
\end{equation}
on condition that $\alpha >2(\beta -1)$ which provides a decrease in the right side of \eqref{GrindEQ__38_}.
Differentiation \eqref{GrindEQ__38_} with respect to $E$ results in relations for the frequencies corresponding to the
action and their derivatives:
\begin{equation} \label{GrindEQ__39_}
\omega E)-\Omega _{big} (E)\sim \frac{1}{E^{3/2+(1-\beta )/\alpha } },\qquad
\Dpi\omega E -\Dpi{\Omega _{big}}E\sim \frac{1}{E^{5/2+(1-\beta )/\alpha } }.
\end{equation}

The difference between the phases $\varphi $ and $\phi $ is written in the form:
$$|\varphi -\phi|\le |\omega -\Omega _{big}|t_{big} +\omega |t-t_{big} |\sim
\frac{t_{big} }{E^{3/2+(1-\beta )/\alpha } } +\omega |\omega -\Omega _{big} |.$$
For high energy frequency $\omega \sim t_{big} \sim \sqrt{E} $ then from here  and \eqref{GrindEQ__39_} we have
\begin{equation} \label{GrindEQ__40_}
\varphi -\phi \sim \frac{1}{E^{1+(1-\beta )/\alpha } } \to 0.
\end{equation}

Let us transform from phase to phase in the integral \eqref{GrindEQ__5_} with accuracy \eqref{GrindEQ__40_} and use the
equality $q(\phi +\pi ,J_{big} )=Q_{2} -q(\phi ,J_{big} )$. Since $2=Q_{2} +Q_{1} $ we rewrite \eqref{GrindEQ__5_} in
the following form:
$$H_{n}=\frac{1}{2}\int\limits_{Q_{1}}^{Q_{2} }dq\left[U^{(1)} (2-q)+U^{(1)} (Q_{1} +q)(-1)^{n}\right]\cos\frac{\pi n}{Q_{2} -Q_{1} }q .$$
A leading term with respect to $E$ is determined by means of integration:
\begin{multline}\label{GrindEQ__54_}
H_{n}=\frac12\left.\left[-U(L-q)+U(Q_{1} +q)(-1)^{n}\right]\cos\frac{\pi n}{Q_{2}-Q_{1} }q\right|_{Q_{1} }^{Q_{2} } -\\
-\frac12\frac{\pi }{Q_{2} -Q_{1} } \int\limits_{Q_{1}}^{Q_{2} }dq\left[U(L-q)-U(Q_{1} +q)(-1)^{n} \right]
\sin\frac{\pi n}{Q_{2}-Q_{1}}q.
\end{multline}
The outside terms have a singular behavior with respect to $E$ since
$$
\begin{aligned}
\frac{(-1)^{n}}2\left[-U(Q_{1} )-U(2Q_{1} )\right]=&
-\frac{(-1)^{n} }2\frac{1}{Q_{1}^{\alpha } }[1+1/2^{\alpha } ]+\underline{O}\left(\frac{1}{Q_{1}^{\beta } }
\right)=\\
&-E\frac{(-1)^{n} }2[1+1/2^{\alpha } ]+\underline{O}\left(\frac{1}{Q_{1}^{\beta } } \right).
\end{aligned}
$$
The integral in \eqref{GrindEQ__54_} is smaller than these terms then the Fourier coefficient is equal to
\begin{equation} \label{GrindEQ__55_}
H_{n} =-E\frac{(-1)^{n} }2 [1+1/2^{\alpha } ]+\ldots
\end{equation}
From \eqref{GrindEQ__9_}, \eqref{GrindEQ__55_} we obtain expression for functions $K_{n,n-1} $ :
$$K_{n,n-1} =\sqrt{\frac{\alpha _{1} }{2} (1+1/2^{\alpha } )} (n-1/2).$$
The parameter $\alpha =12$ for the Lennard-Jones potential then we obtain\break
$K_{2,1} =1.5\sqrt{\alpha _{1} /2} $, $K_{3,2} =2.5\sqrt{\alpha _{1} /2}$.

\subsection{Comparison with numerical results for $K_{2,1}(q_0)$}

The function $K_{2,1} (q_{0} )$ was defined numerically (exact) and calculated (asympt) in accordance with formula
\eqref{GrindEQ__17_}. Its behavior depending on the initial position of the resonance trajectory is presented in Fig.1
(exact). From here it is seen that formula \eqref{GrindEQ__17_} is  a good approximation for $K_{2,1} (q_{0} )$ as for
small as high energies.

\dA=\hsize\hsize=12cm\dB=\dA\advance\dB-\hsize\divide\dB2
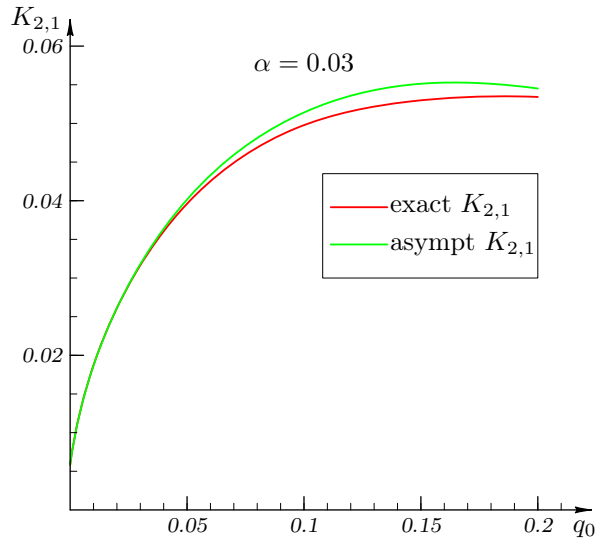
\setbox\wbox=\vbox to 0pt{\small\noindent Fig. 1. Graph of the function $K_{2,1}(q_{0})$
defined numerically (exact) and \centerline{\hbox to30pt{}calculated (asympt) in accordance with
formula \eqref{GrindEQ__17_}.}\vss}
\hsize=\dA
\ifpdf
    \centerline{\includegraphics{k21-1}}
\else
    \centerline{\includegraphics{k21.1}}
\fi

\noindent\mbox{\kern\dB}\box\wbox
\vskip3\baselineskip
\subsection{Acknoledgments}
This work was supported by Russian Foundation for Basic Research (Project ¹11-01-12057-ofi-i-2011)

\subsection*{\indent References}
\begin{enumerate}\parsep=0pt\parskip=0pt\itemsep=0pt

\item{{\sl Zaslavsky G.\:M., Sagdeev R.\:Z., Usikov D.\:A., Chernikov A.\:A.} Weak chaos and quasi-regular patterns.
Cambridge University Press, 1991. 254p.}

\item{{\sl Guzev M.\:A., Koshel K.\:V., Izrailsky Yu.\:G.} The effect of global chaos in a chain of particles//
Nonlinear Dynamics. 2010. V.~52, No.~5, pp.~291--305 (in Russian)(http://nd.ics.org.ru/doc/r/pdf/1676/0)}

\item{{\sl Chirikov B.V.} A universal instability of many-dimensional oscillator systems// Phys. Rep., 1979. V.~52, No.~5, pp.~264--379.}

\end{enumerate}

\end{document}